\documentclass[aps,prb,reprint,superscriptaddress,showpacs,showkeys,twocolumn]{revtex4} 
\usepackage{amsfonts}
\usepackage{amssymb}
\usepackage{amsmath}
\usepackage{graphicx}
\usepackage{color}

\begin{document} 

\title{Tunable asymmetric magnetoimpedance effect in ferromagnetic NiFe/Cu/Co films} 
\author{E.~F.~Silva} 
\affiliation{Departamento de F\'{i}sica Te\'{o}rica e Experimental, Universidade Federal do Rio Grande do Norte, 59078-900 Natal, RN, Brazil}
\author{M.~Gamino}
\affiliation{Instituto de F\'{i}sica, Universidade Federal do Rio Grande de Sul, 91501-970 Porto Alegre, RS, Brazil}
\author{A.~M.~H.~de Andrade}
\affiliation{Instituto de F\'{i}sica, Universidade Federal do Rio Grande de Sul, 91501-970 Porto Alegre, RS, Brazil}
\author{M.~A.~Corr\^{e}a}
\affiliation{Departamento de F\'{i}sica Te\'{o}rica e Experimental, Universidade Federal do Rio Grande do Norte, 59078-900 Natal, RN, Brazil} 
\author{M.~V\'{a}zquez}
\affiliation{Instituto de Ciencia de Materiales de Madrid, CSIC, 28049 Madrid, Spain} 
\author{F.~Bohn} 
\email[Electronic address: ]{felipebohn@gmail.com}
\affiliation{Departamento de F\'{i}sica Te\'{o}rica e Experimental, Universidade Federal do Rio Grande do Norte, 59078-900 Natal, RN, Brazil} 

\date{\today} 

\begin{abstract} 
We investigate the magnetization dynamics through the magnetoimpedance effect in ferromagnetic NiFe/Cu/Co films. We observe that the magnetoimpedance response is dependent on the thickness of the non-magnetic Cu spacer material, a fact associated to the kind of the magnetic interaction between the ferromagnetic layers. Thus, we present an experimental study on asymmetric magnetoimpedance in ferromagnetic films with biphase magnetic behavior and explore the possibility of tuning the linear region of the magnetoimpedance curves around zero magnetic field by varying the thickness of the non-magnetic spacer material, and probe current frequency. We discuss the experimental magnetoimpedance results in terms of the different mechanisms governing the magnetization dynamics at distinct frequency ranges, quasi-static magnetic properties, thickness of the non-magnetic spacer material, and the kind of the magnetic interaction between the ferromagnetic layers. The results place ferromagnetic films with biphase magnetic behavior exhibiting asymmetric magnetoimpedance effect as a very attractive candidate for application as probe element in the development of auto-biased linear magnetic field sensors.
\end{abstract}



\maketitle 
The magnetoimpedance effect (MI), known as the change of the real and imaginary components of electrical impedance of a ferromagnetic conductor caused by the action of an external static magnetic field, is commonly employed as a tool to investigate ferromagnetic materials. For a general review on the effect, we suggest the Ref.~\cite{Resumao_MI}. In recent years, the interest for this phenomenon has grown considerably not only for its contribution to the understanding of fundamental physics associated to magnetization dynamics~\cite{APL69p3084}, but also due to the possibility of application of materials exhibiting magnetoimpedance as probe element in sensor devices for low-field detection~\cite{JMMM293p671}. In this sense, the sensitivity and linearity as a function of the magnetic field are the most important parameters in the practical application of magnetoimpedance effect for magnetic sensors~\cite{APL75p2114}. Experiments have been carried out in numerous magnetic systems, including ribbons~\cite{PRB53pR5982, PRB60p6685, JAP95p1364}, sheets~\cite{JAP84p3792}, wires~\cite{APL64p3652, PRB50p16737, APL77p121, SAA59p20, JAP87p4822}, and, magnetic films~\cite{APL67p3346, JAP101p033908, APL94p042501, JPDAP43p295004, TSF520p2173, JMMM242p291, JPCM16p6561, JPDAP41p175003, JPDAP43p295004, APL94p042501, JAP115p103908, APL104p102405}. However, although soft magnetic materials are highly sensitive to small field variations at low magnetic fields, due to magnetization process most of them essentially have nonlinear MI behavior around zero magnetic field, which prevents a simple straightforward derivation of an appropriate signal for sensor applications~\cite{APL104p102405, JAP105p033911}. 

The shift of the sensor operational region and the leading of the linear MI behavior at around zero magnetic field can be obtained primarily by applying a bias field or an electrical current to the ordinary MI element~\cite{JAP105p033911}. However, this approach proved to be disadvantageous from the practical and technological point of view, mainly due to energetic consumption. Recently, it has been shown that materials exhibiting asymmetric magnetoimpedance (AMI) effect arise as promising alternative with potential of application, opening possibilities for the use of this kind of materials for the development of auto-biased linear magnetic field sensors. For these materials, the asymmetric effects are obtained by inducing an asymmetric static magnetic configuration, usually done by magnetostatic interactions~\cite{JAP105p033911, JMMM295p121, JAP87p4822} or exchange bias~\cite{APL85p3507, APL94p042501, APL96p232501, APL104p102405, APL75p2114}.

For ferromagnetic films, the primary AMI results have been measured for exchange biased multilayers~\cite{APL94p042501, APL96p232501, APL104p102405}. Theory and experiment agree well for MI curves shifted by the exchange bias field, following the main features of the magnetization curve, as well as it is verified that the linear region of AMI curves can be tuned to around zero just by modifying the angle between applied magnetic field and exchange bias field, or changing the probe current frequency. On the other hand, another promising possibility of AMI material resides in films presenting biphase magnetic behavior, with hard and soft ferromagnetic phases intermediated by a non-magnetic layer acting together.

In this work, we investigate the magnetoimpedance effect in ferromagnetic NiFe/Cu/Co films. We observe that the MI response is dependent on the thickness of the non-magnetic Cu spacer material, a fact associated to the kind of the magnetic interaction between the ferromagnetic layers. Here we show that the linear region of the asymmetric magnetoimpedance curves in these films is experimentally tunable by varying the thickness of the non-magnetic spacer material, and probe current frequency. The results place ferromagnetic films with biphase magnetic behavior exhibiting asymmetric magnetoimpedance effect as a very attractive candidate for application as probe element in the development of auto-biased linear magnetic field sensors.

For this study, we produce Ni$_{81}$Fe$_{19}$($25$ nm)/Cu($t_{Cu}$)/Co($50$ nm) ferromagnetic films, with $t_{Cu} = 0$, $1.5$, $3$, $5$, $7$, and $10$ nm. The films are deposited by magnetron sputtering from targets of nominally identical compositions onto glass substrates, with dimensions of $8$ $\times$ $4$ mm$^2$. A buffer Ta layer is deposited before the NiFe layer to reduce the roughness of the substrate, as well as a cap Ta layer is inserted after the Co layer in order to avoid oxidation of the sample. The deposition is carried out with the following parameters: base vacuum of $10^{-8}$ Torr, deposition pressure of $2.0$ mTorr with a $99.99$\% pure Ar at $32$ sccm constant flow, and DC source with power of $150$ W for the deposition of the NiFe and Co layers, while $100$ W for the Cu and Ta layers. During the deposition, the substrate rotates at constant speed to improve the film uniformity, and a constant magnetic field of $2$ kOe is applied perpendicularly to the main axis of the substrate in order to induce a magnetic anisotropy and define an easy magnetization axis. X-ray diffraction results, not shown here, calibrate the deposition rates and verify the Co$(111)$ and NiFe$(111)$ preferential growth of all films. Magnetization curves are obtained with a vibrating sample magnetometer, measured along and perpendicular to the main axis of the films, to verify the quasi-static magnetic behavior. Magnetization dynamics is investigated through MI measurements obtained using a RF-impedance analyzer Agilent model $E4991$, with $E4991A$ test head connected to a microstrip in which the sample is the central conductor. Longitudinal MI measurements are performed by acquiring the real $R$ and imaginary $X$ parts of the impedance $Z$ over a wide range of frequencies, from $0.1$ GHz up to $3.0$ GHz, with $0$ dBm ($1$ mW) constant power applied to the sample, characterizing the linear regime of driving signal, and magnetic field varying between $\pm 300$ Oe, applied along the main axis of the sample. Detailed information on the MI experiment is found in Refs.~\cite{JPDAP41p175003, JAP115p103908}. In order to quantify the sensitivity and MI performance as a function of the frequency, we calculate the magnitude of the impedance change at the low field range $\pm 6$ Oe using the expression~\cite{APL104p102405}
\begin{equation}
 \frac{|\Delta Z|}{|\Delta H|} = \frac{\left|Z(H=6\,\textrm{Oe}) - Z(H=-6\,\textrm{Oe}) \right|}{12}.
\label{eq_01}
\end{equation}
\noindent Here, we consider the absolute value of $\Delta Z$, since the impedance around zero field can present positive or negative slopes, depending on the sample and measurement frequency. In particular, it is verified that $|\Delta Z|/|\Delta H|$ is roughly constant at least for a reasonable low field range. 

\begin{figure*}[!] 
\hspace{-.25cm}\includegraphics[width=5.85cm]{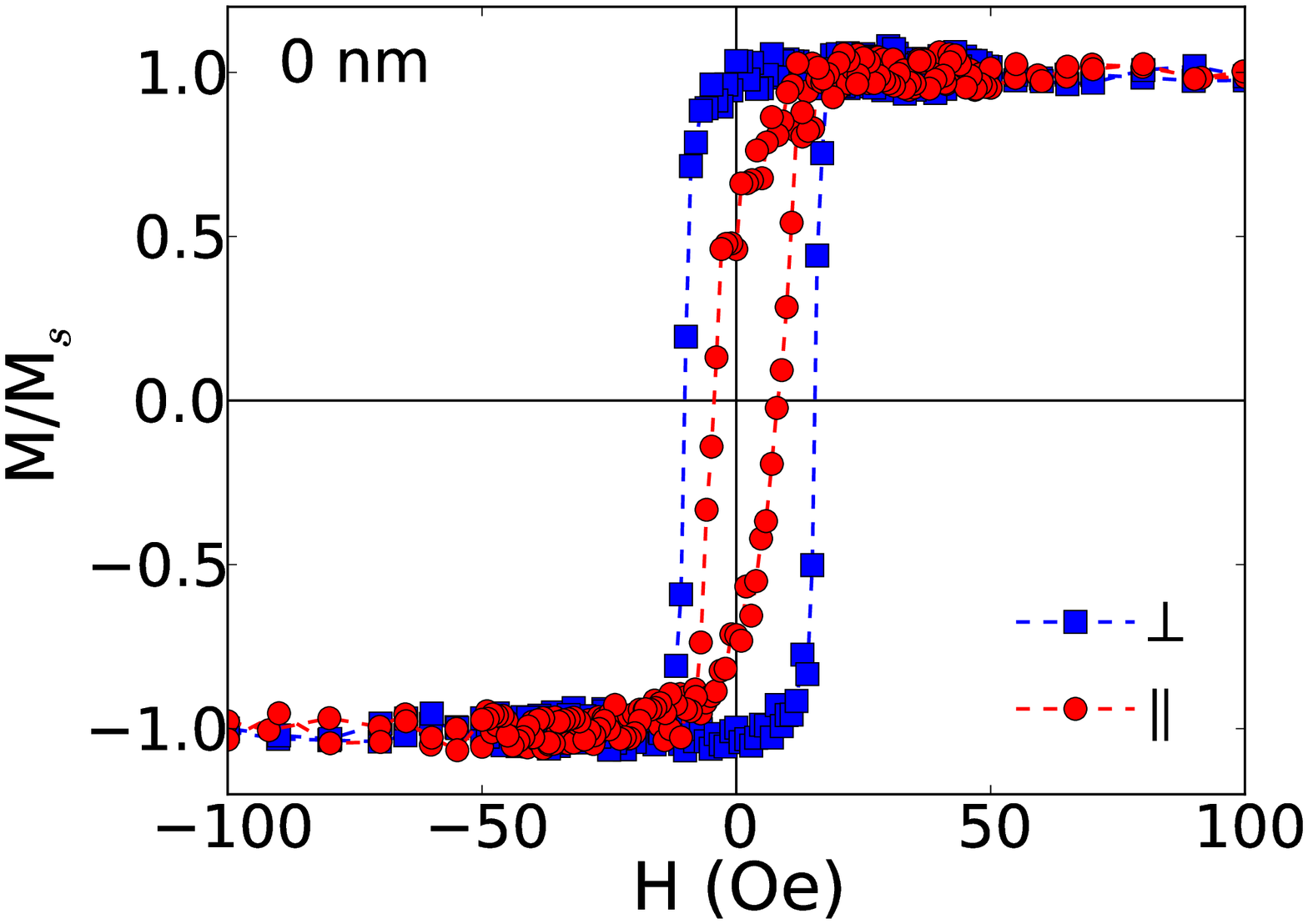} 
\hspace{-.25cm}\includegraphics[width=5.85cm]{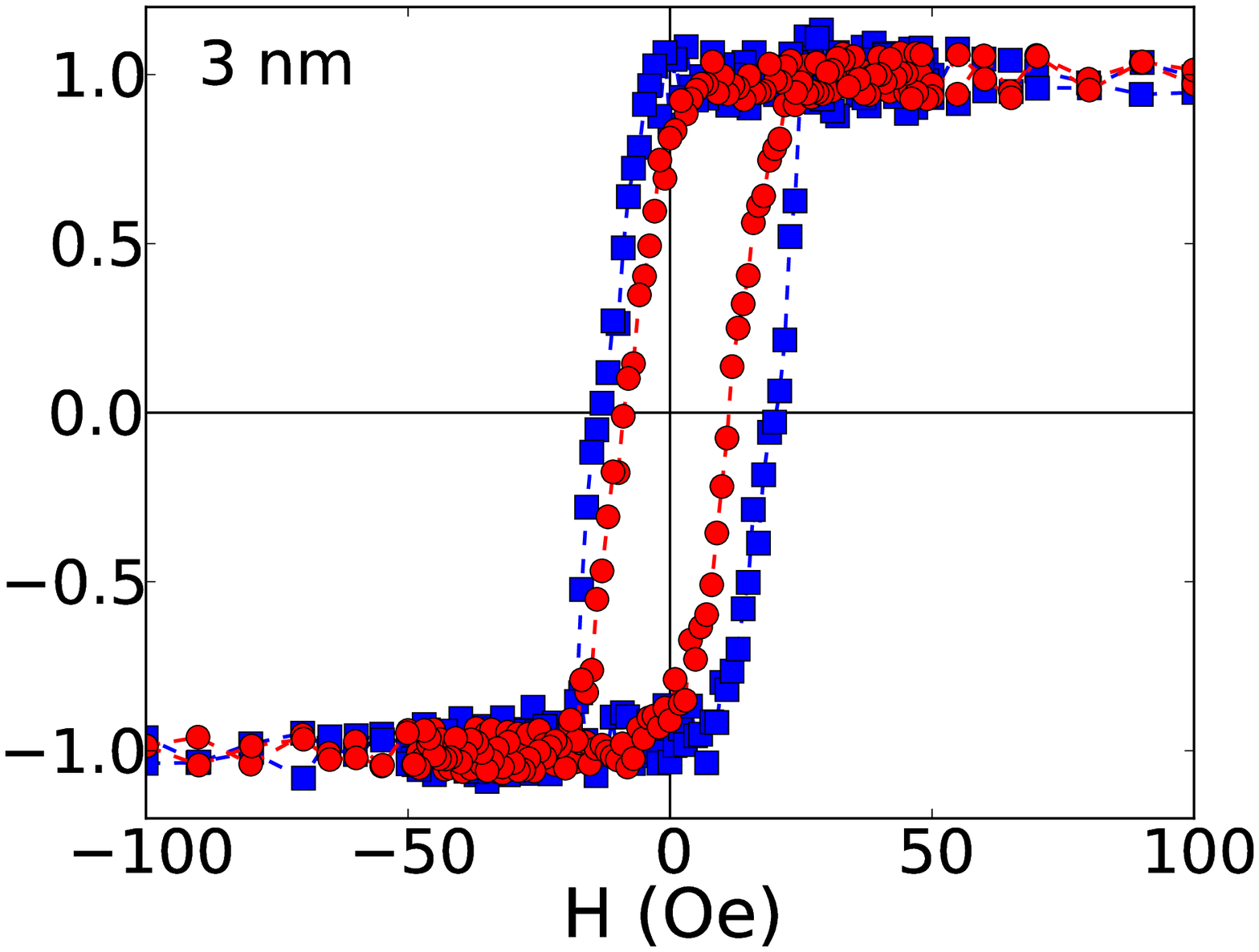}
\hspace{-.25cm}\includegraphics[width=5.85cm]{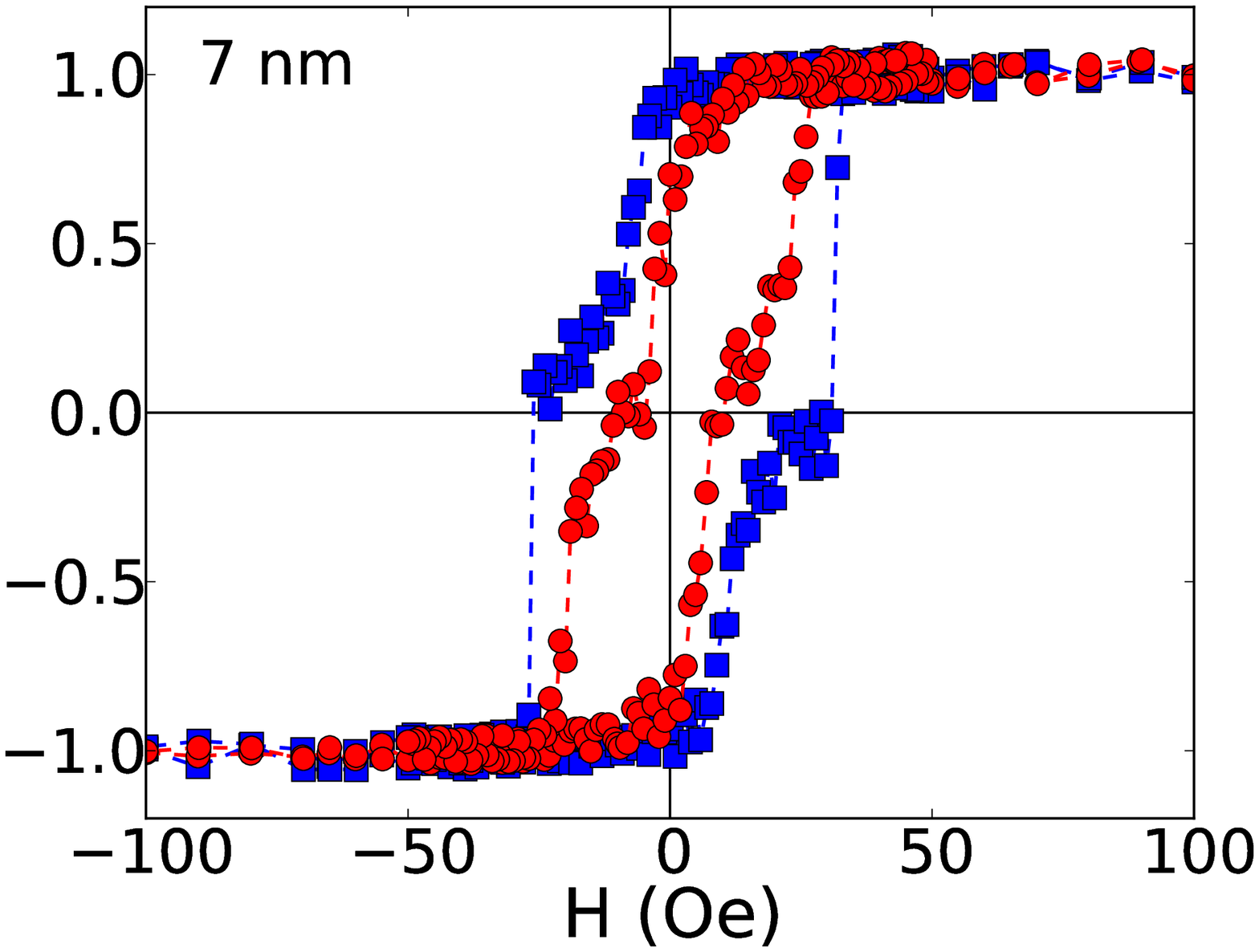}
\vspace{-.3cm}\caption{Representative normalized quasi-static magnetization curves for selected NiFe/Cu/Co films with different thicknesses of the non-magnetic Cu spacer material, obtained with the in-plane magnetic field applied along ($\parallel$) and perpendicular ($\perp$) to the main axis. Films with $t_{Cu}$ below $3$ nm present behavior similar to that verified for the film with $t_{Cu} = 1.5$ nm, while the films with $t_{Cu}$ above the critical thickness range have behavior identical to the one observed for the film with $t_{Cu} = 7$ nm. The film with $t_{Cu}= 3$ nm is within the critical Cu thickness range and have an intermediate behavior.} 
    \label{Fig_01} 
\end{figure*}

Figure~\ref{Fig_01} shows the quasi-static magnetization curves for selected films, measured with the external in-plane magnetic field applied along and perpendicular to the main axis. When analyzed as a function of the the thickness of the non-magnetic Cu spacer material, it is observed an evolution of the shape of the magnetization curves, indicating the existence of a critical thickness range, $\sim 3$ nm, which splits the films in groups according the magnetic behavior. For films with $t_{Cu}$ below $3$ nm, the NiFe and Co layers are ferromagnetically coupled. The angular dependence of the magnetization curves indicates an uniaxial in-plane magnetic anisotropy, induced by the magnetic field applied during the deposition process, and oriented perpendicularly to the main axis. Despite the similar magnetic behavior, the film with $1.5$ nm-thick Cu layer (not shown) has slightly higher coercive field if compared to the one for the film without spacer material, possibly associated to the increase of the whole sample disorder due to the non-formation of a regular complete Cu layer. The film with $t_{Cu}= 3$ nm, within the critical Cu thickness range, presents an intermediate magnetic behavior, with smaller magnetic permeability, characterized by the first evidences of a small plateau in the measurement perpendicular to the main axis, and the appearance of magnetization regions associated to distinct anisotropy constants of the NiFe and Co layers. Films with $t_{Cu}$ above the critical thickness range exhibit a biphase magnetic behavior. The two-stage magnetization process is characterized by the magnetization reversion of the soft NiFe layer at low magnetic field, followed by the reversion of the hard Co layer at higher field. None substantial difference between the magnetization curves measured for films with $t_{Cu} > 3$ nm is verified. In principle, the biphase magnetic behavior suggests that the ferromagnetic layers are uncoupled. The easy magnetization axis remains perpendicular to the main axis of the substrate, as expected. The weaker anisotropy induction and increase of hysteretic losses are primarily related to the roughness of the interfaces and lack of homogeneity of the Cu layer arisen as its thickness is raised~\cite{TSF520p2173}. 

It is well-known that quasi-static magnetic properties play a fundamental role in the dynamic magnetic response and are reflected in the MI behavior~\cite{APL104p102405}. The shape and amplitude of the magnetoimpedance curves are strongly dependent on the orientation of the applied magnetic field and {\it ac} current with respect to the magnetic anisotropies, magnitude of the external magnetic field, and probe current frequency, as well as are directly related to the main mechanisms responsible for the transverse magnetic permeability changes: skin effect and ferromagnetic resonance (FMR) effect~\cite{APL67p857, JAP110p093914, APL104p102405}. However, magnetoimpedance effect can also provide a further insights on the nature of the interactions governing the magnetization dynamics and energy terms affecting the transverse magnetic permeability.

\begin{figure}[!] 
\begin{center}
\includegraphics[width=4.35cm]{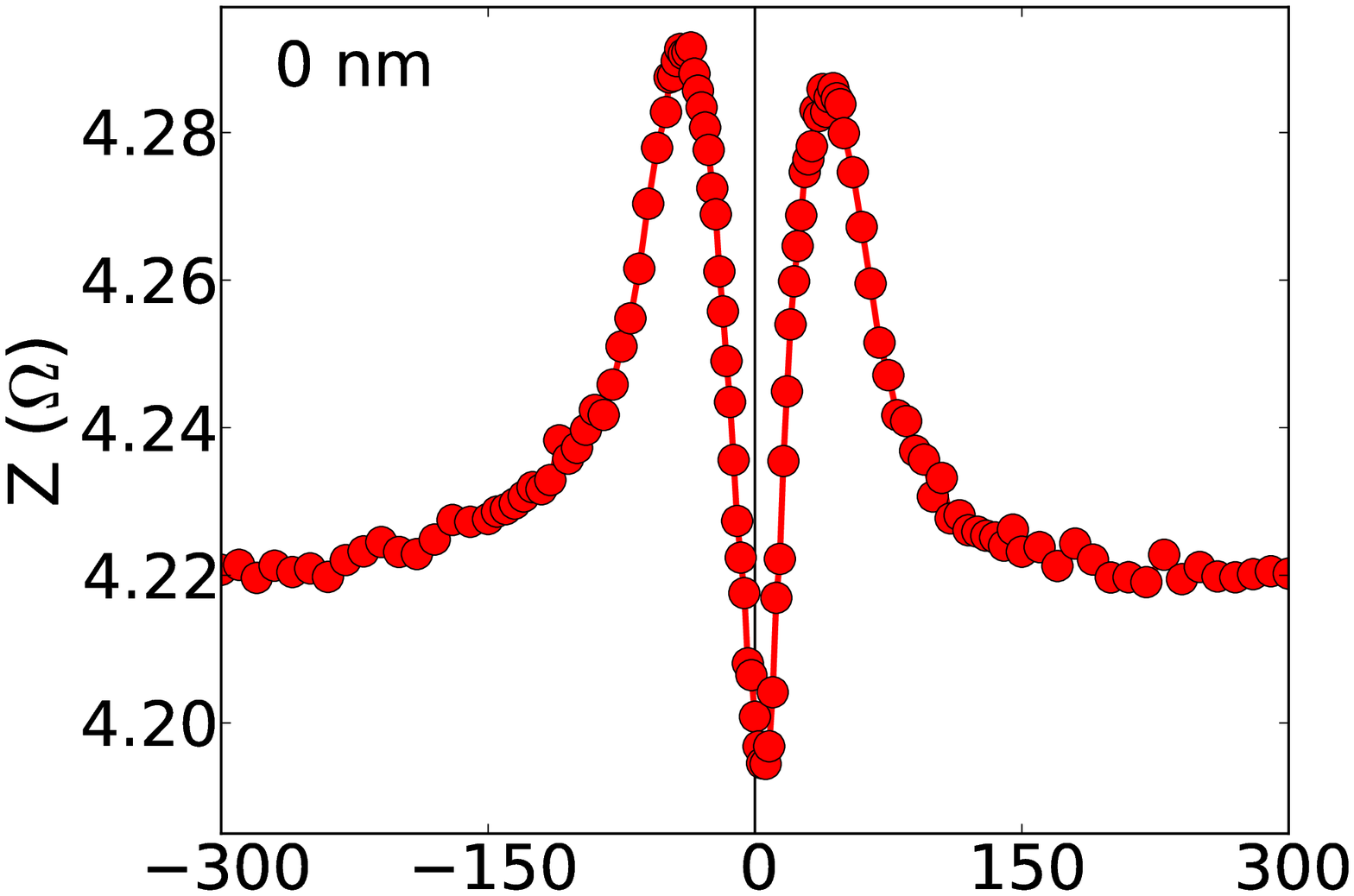} 
\hspace{-.3cm}\includegraphics[width=4.35cm]{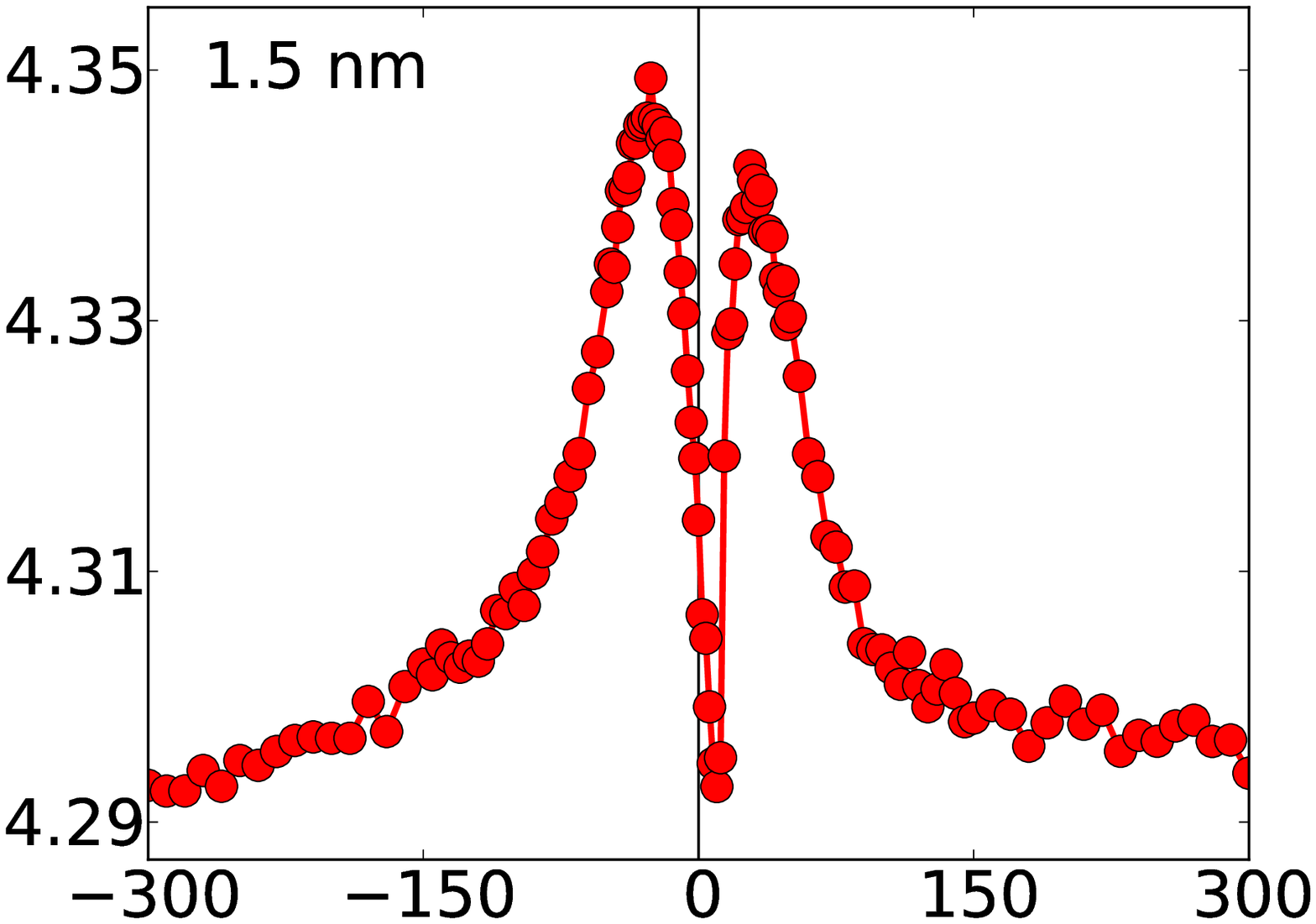}\\
\includegraphics[width=4.35cm]{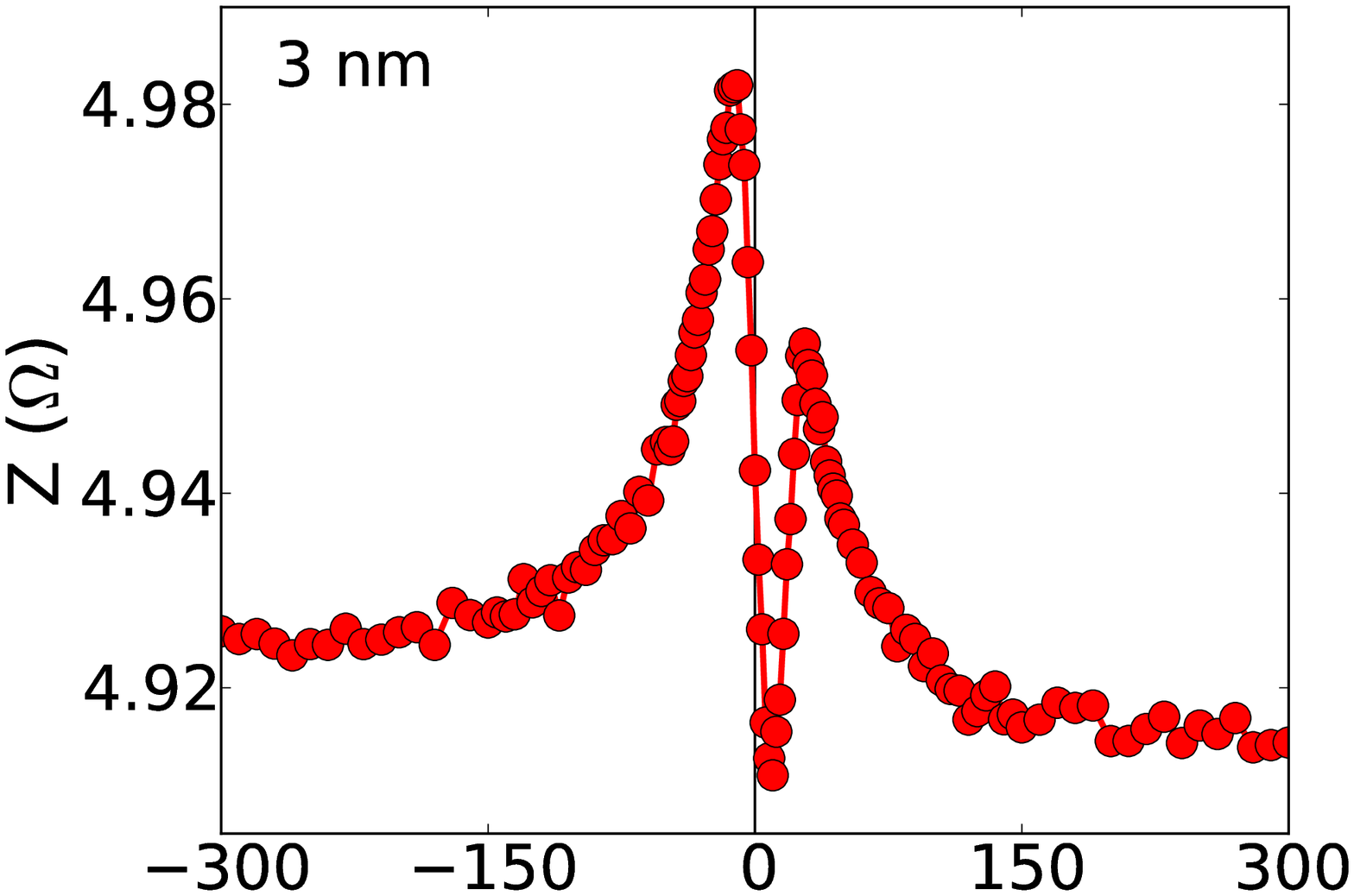}
\hspace{-.3cm}\includegraphics[width=4.35cm]{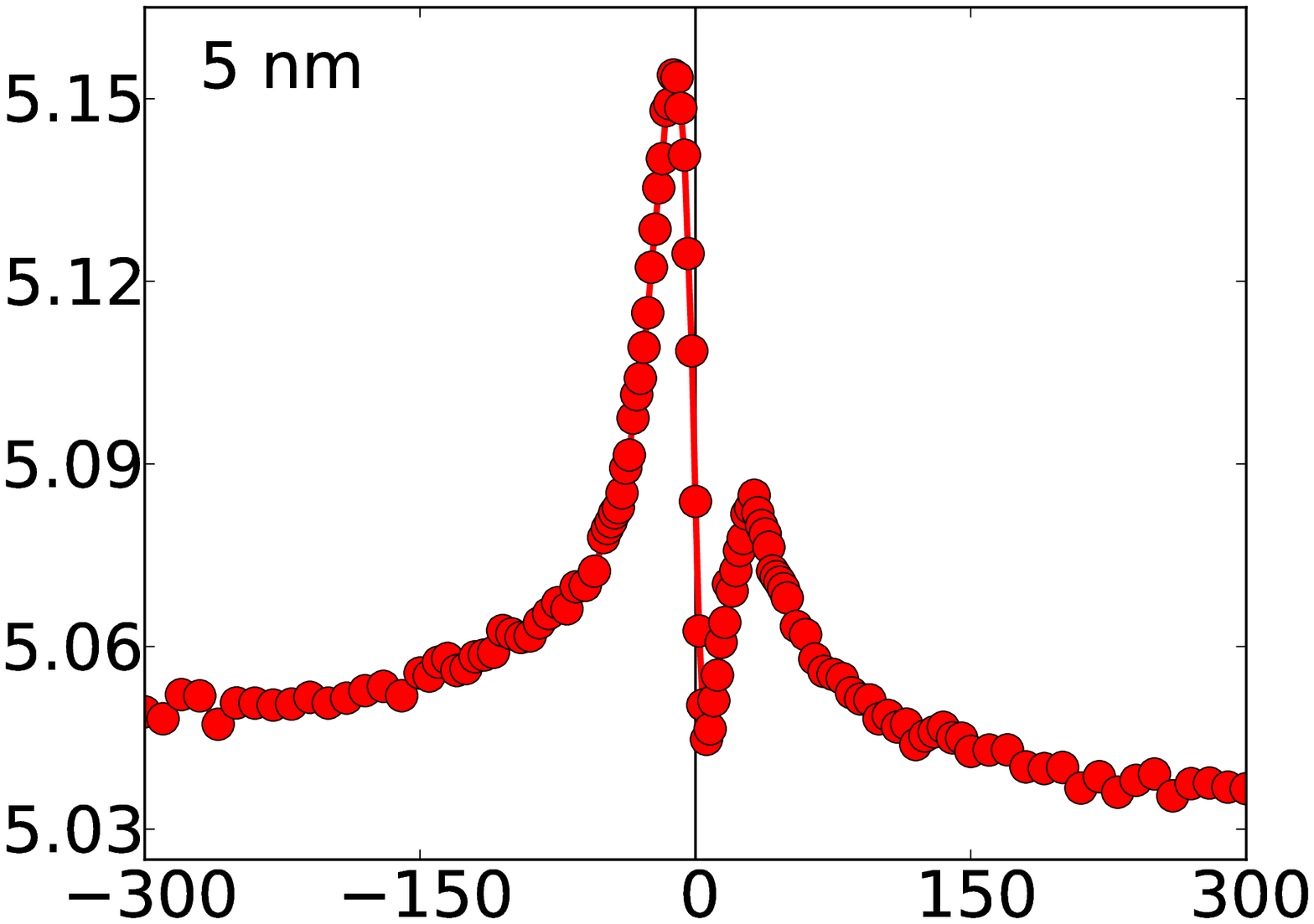}\\ 
\includegraphics[width=4.35cm]{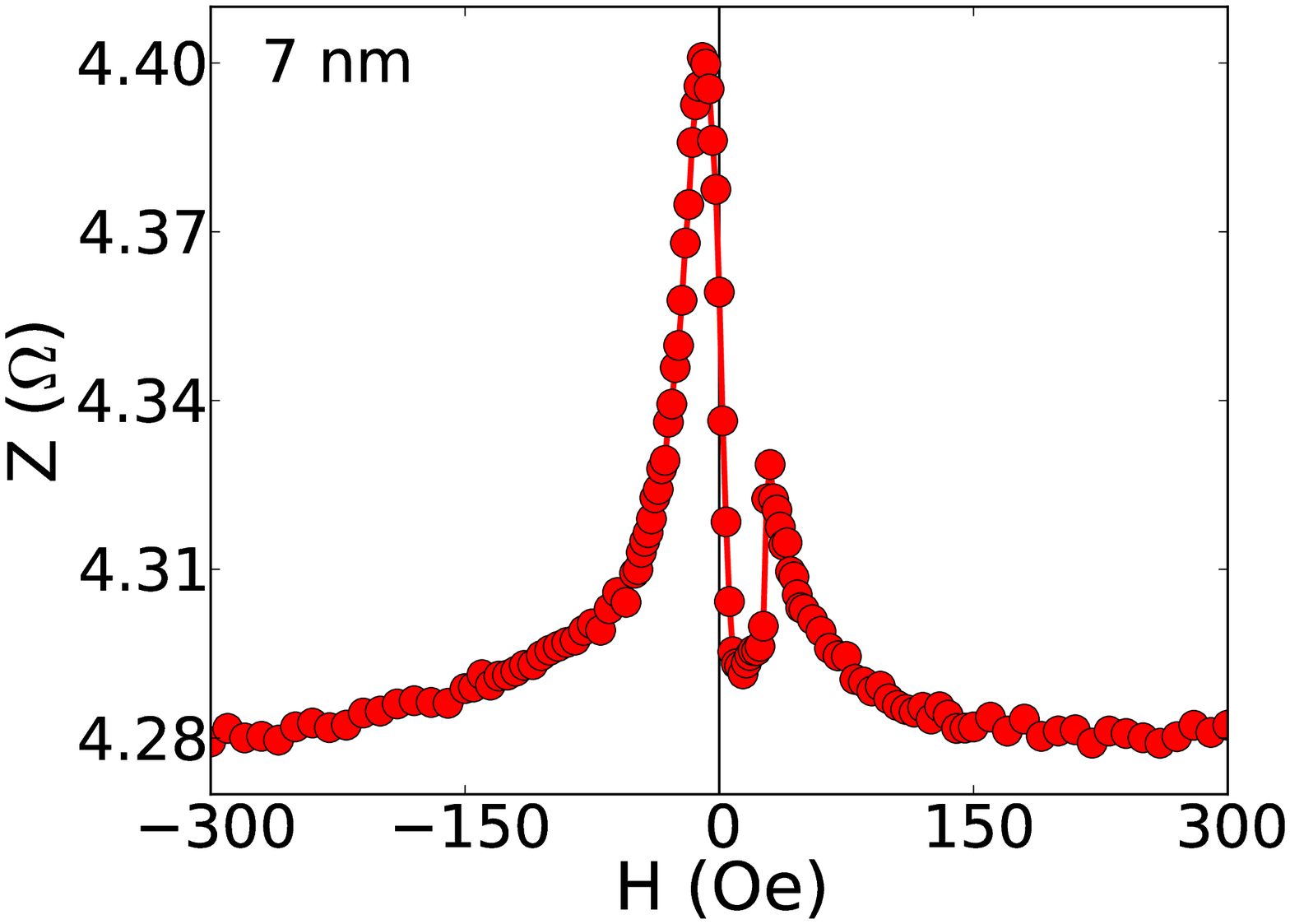}
\hspace{-.3cm}\includegraphics[width=4.35cm]{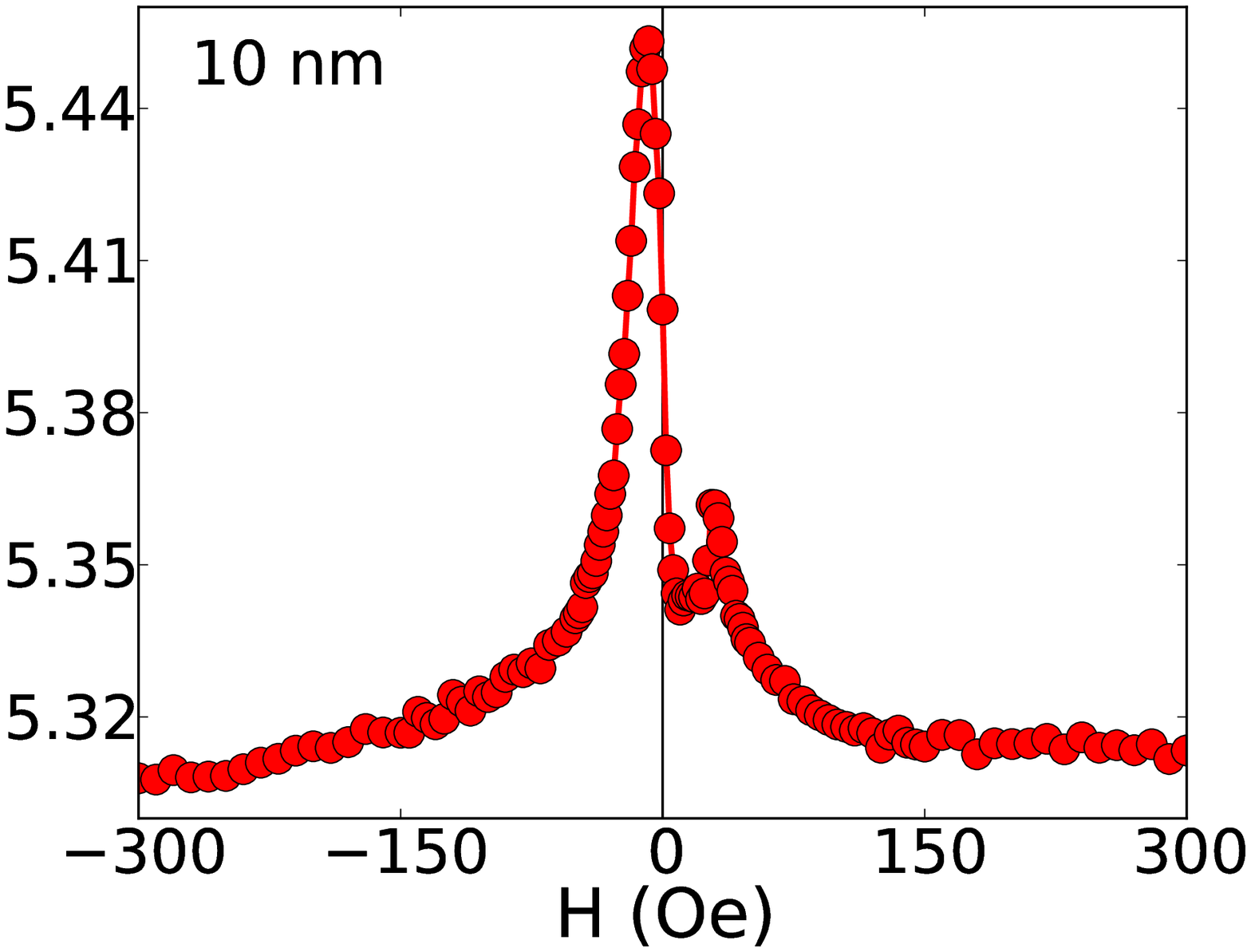}
\end{center}
\vspace{-.3cm}\caption{The MI curves at frequency of $0.75$ GHz measured for the films with different thicknesses of the non-magnetic Cu spacer material $t_{Cu}$. The MI curves are acquired over a complete magnetization loop and present hysteretic behavior. Here, we show just part of the curve, when the field goes from negative to positive values, to make easier the visualization of the whole MI behavior.
} 
\label{Fig_02} 
\end{figure}

Regarding the MI results, Fig.~\ref{Fig_02} shows the MI curves, at the selected frequency of $0.75$ GHz, for the films with different thicknesses $t_{Cu}$ of the non-magnetic Cu spacer material. All samples exhibit a double peak behavior for the whole frequency range, a signature of the perpendicular alignment of the external magnetic field and {\it ac} current with the easy magnetization axis. An interesting feature related to the MI behavior of the NiFe/Cu/Co films resides in the amplitude and position of the peaks with the thickness of the non-magnetic Cu spacer material, and the probe current frequency. 

Films with $t_{Cu} < 3$ nm present the well-known symmetric magnetoimpedance behavior for anisotropic systems. The MI curves have the double peak behavior, symmetrical at aroud $H = 0$, with peaks with roughly the same amplitude. For frequencies up to $\sim 0.85$ GHz, the position of the peaks remains unchanged close to the anisotropy field, indicating that the skin effect is the main responsible by the changes of transverse magnetic permeability governing the magnetization dynamics. For frequencies above this value, not presented here, besides the skin effect, the FMR effect also becomes an important mechanism responsible for variations of the MI effect, a fact evidenced by the displacement of the peaks position toward higher fields as the frequency is increased. The contribution of the FMR effect to $Z$ is also verified using the method described by Barandiar\'{a}n {\it et al.}~\cite{JAP99p103904}, and previously employed by our group~\cite{JPDAP43p295004}. 

\begin{figure}[!] 
\begin{center}
\includegraphics[width=4.35cm]{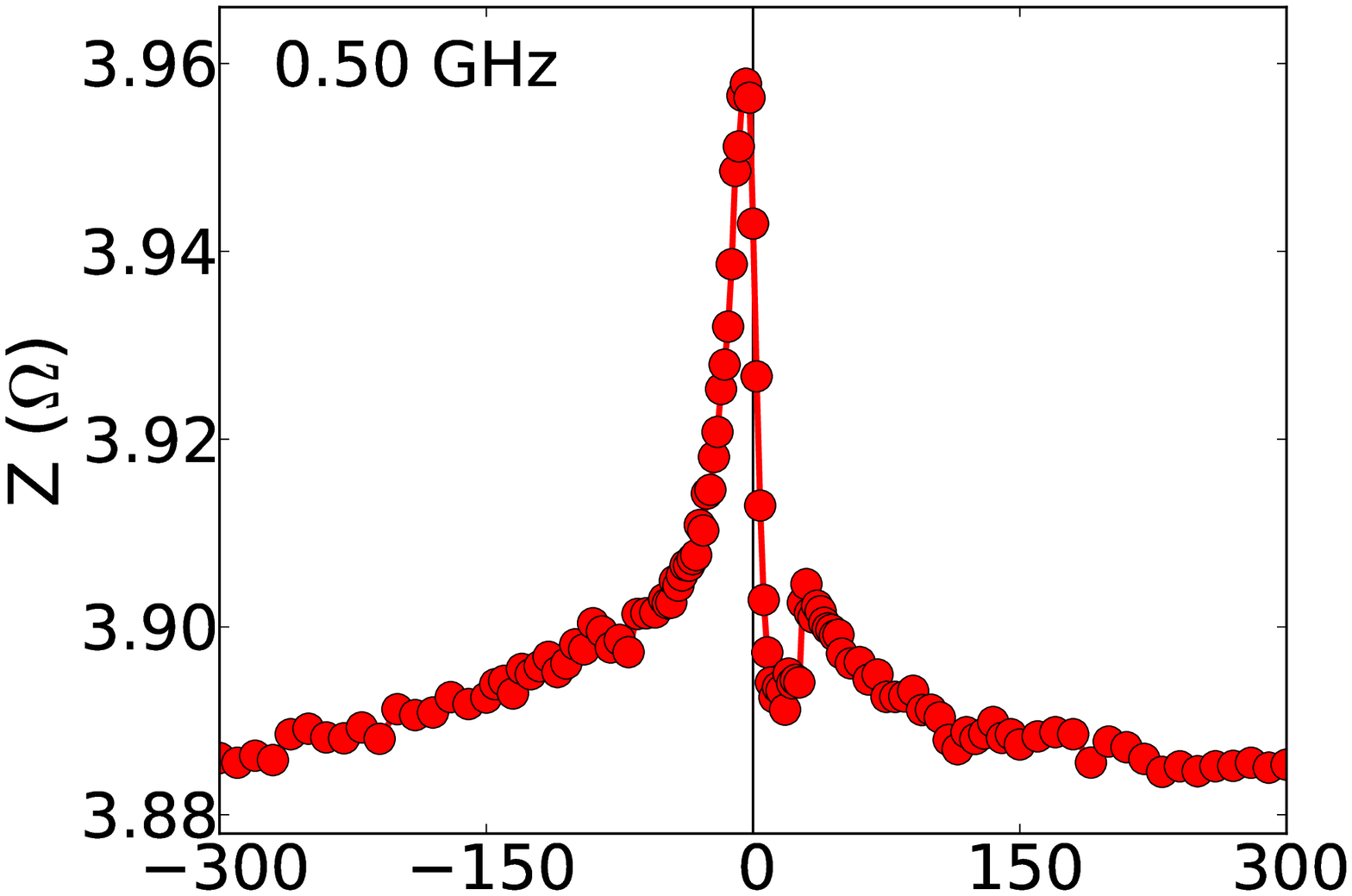} 
\hspace{-.3cm}\includegraphics[width=4.35cm]{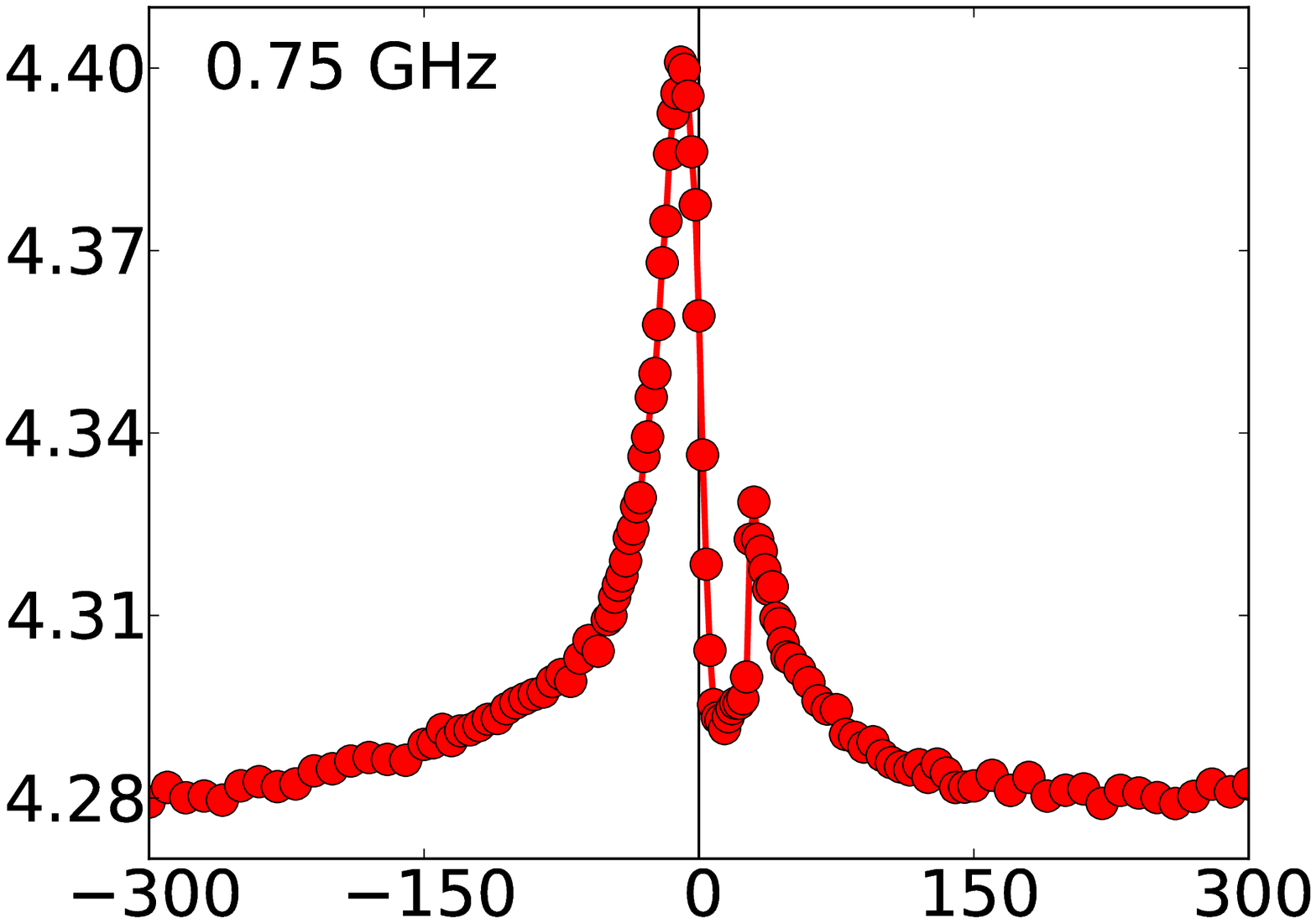}\\
\includegraphics[width=4.35cm]{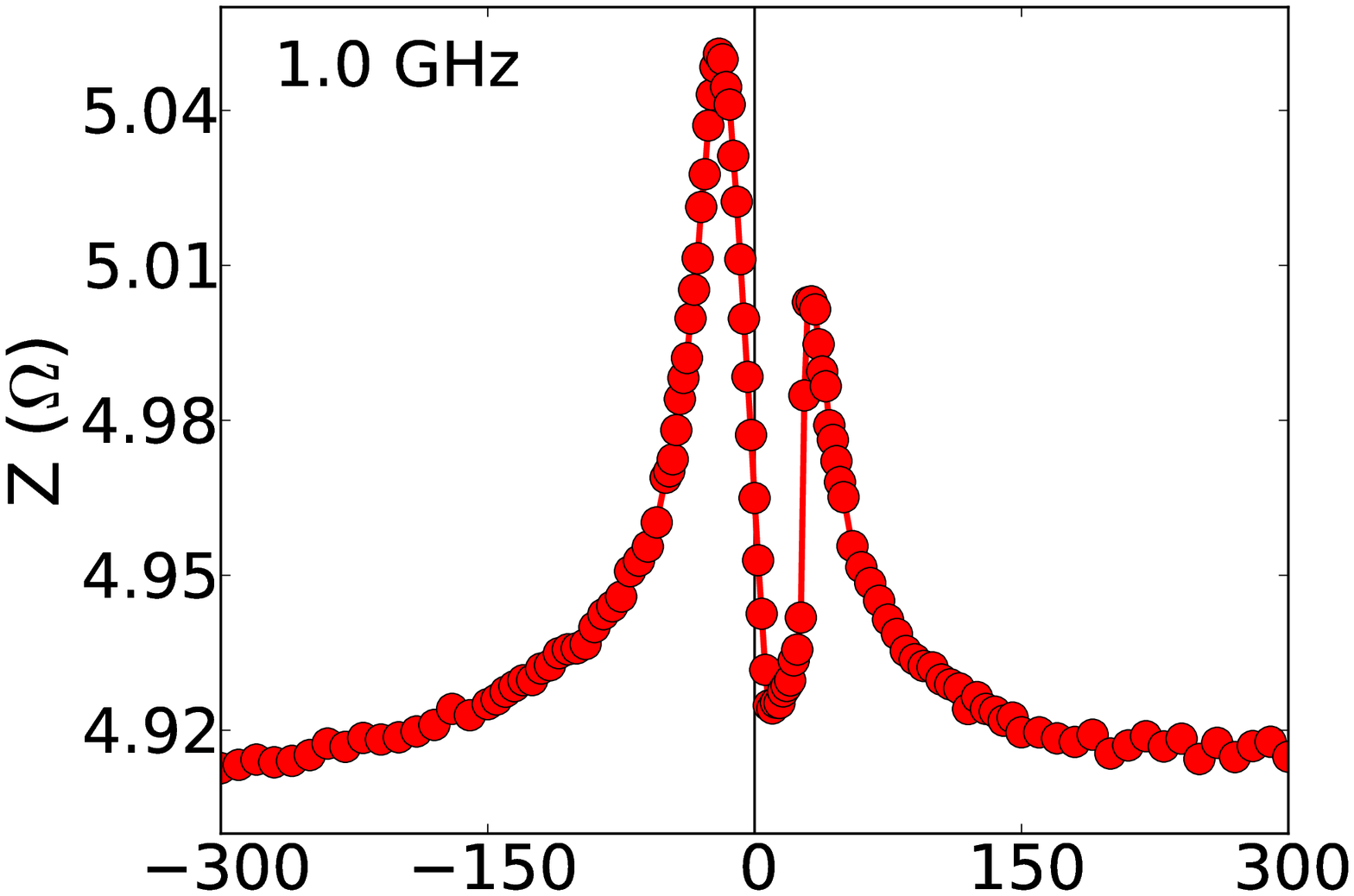}
\hspace{-.3cm}\includegraphics[width=4.35cm]{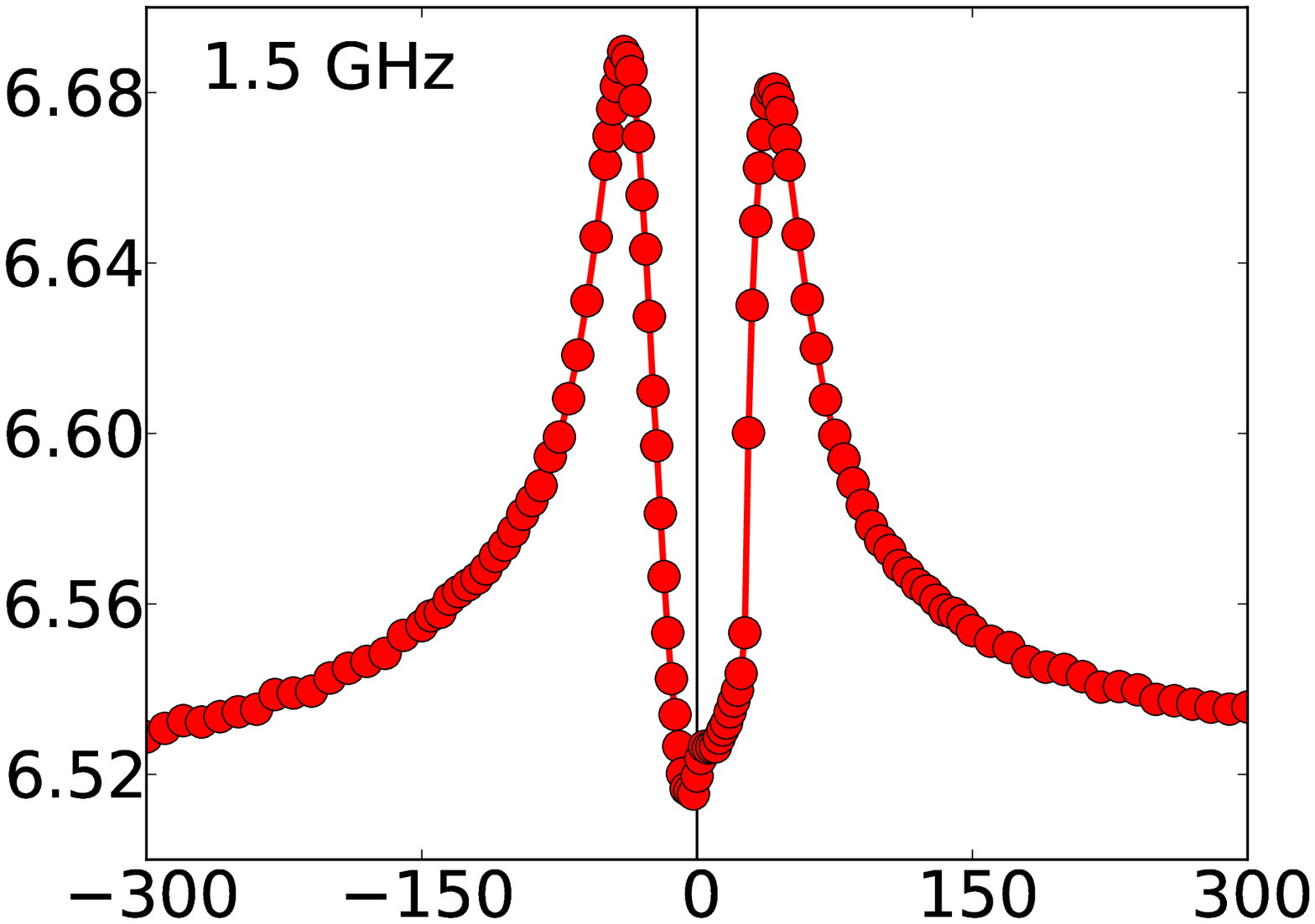}\\ 
\includegraphics[width=4.35cm]{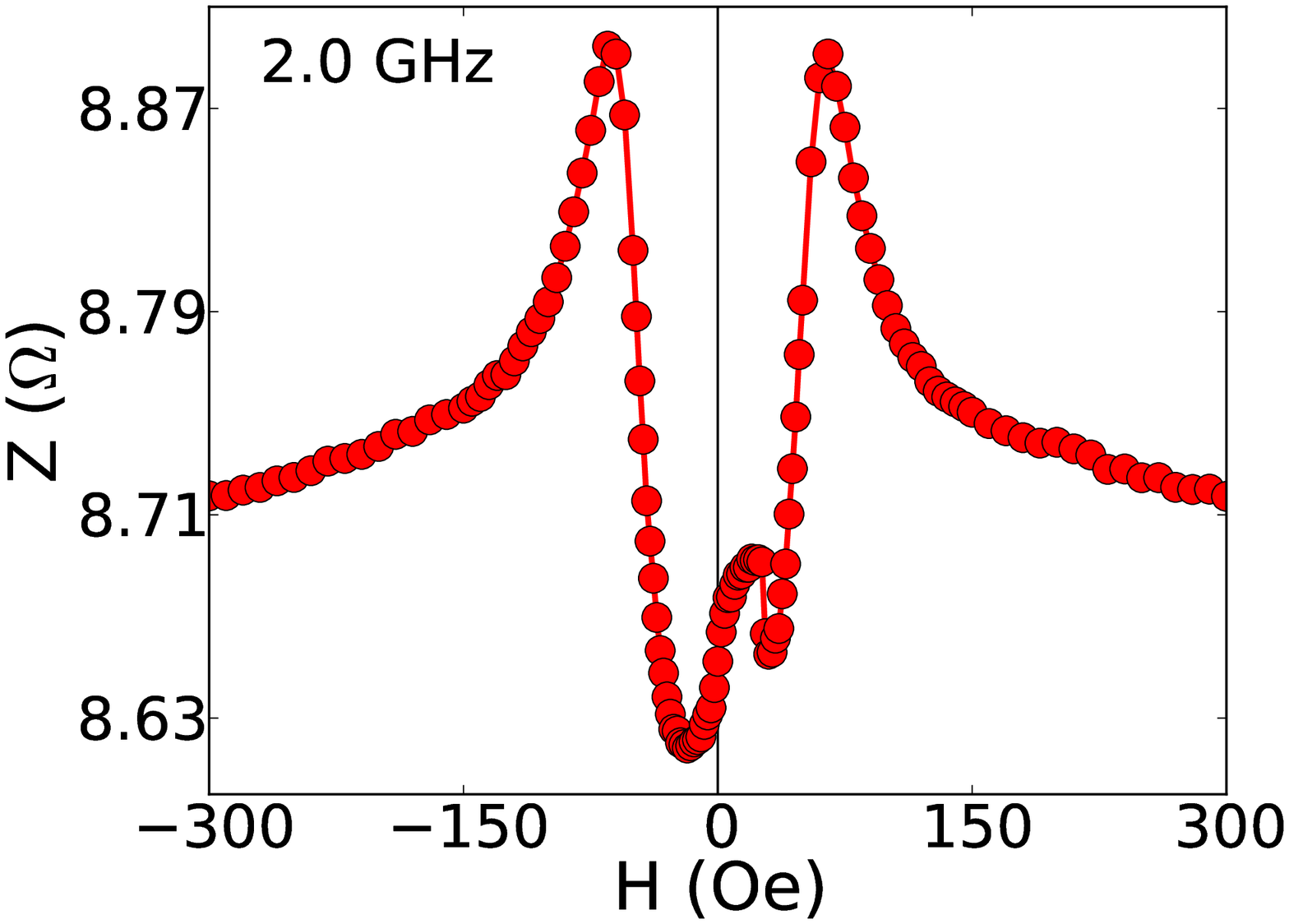}
\hspace{-.3cm}\includegraphics[width=4.35cm]{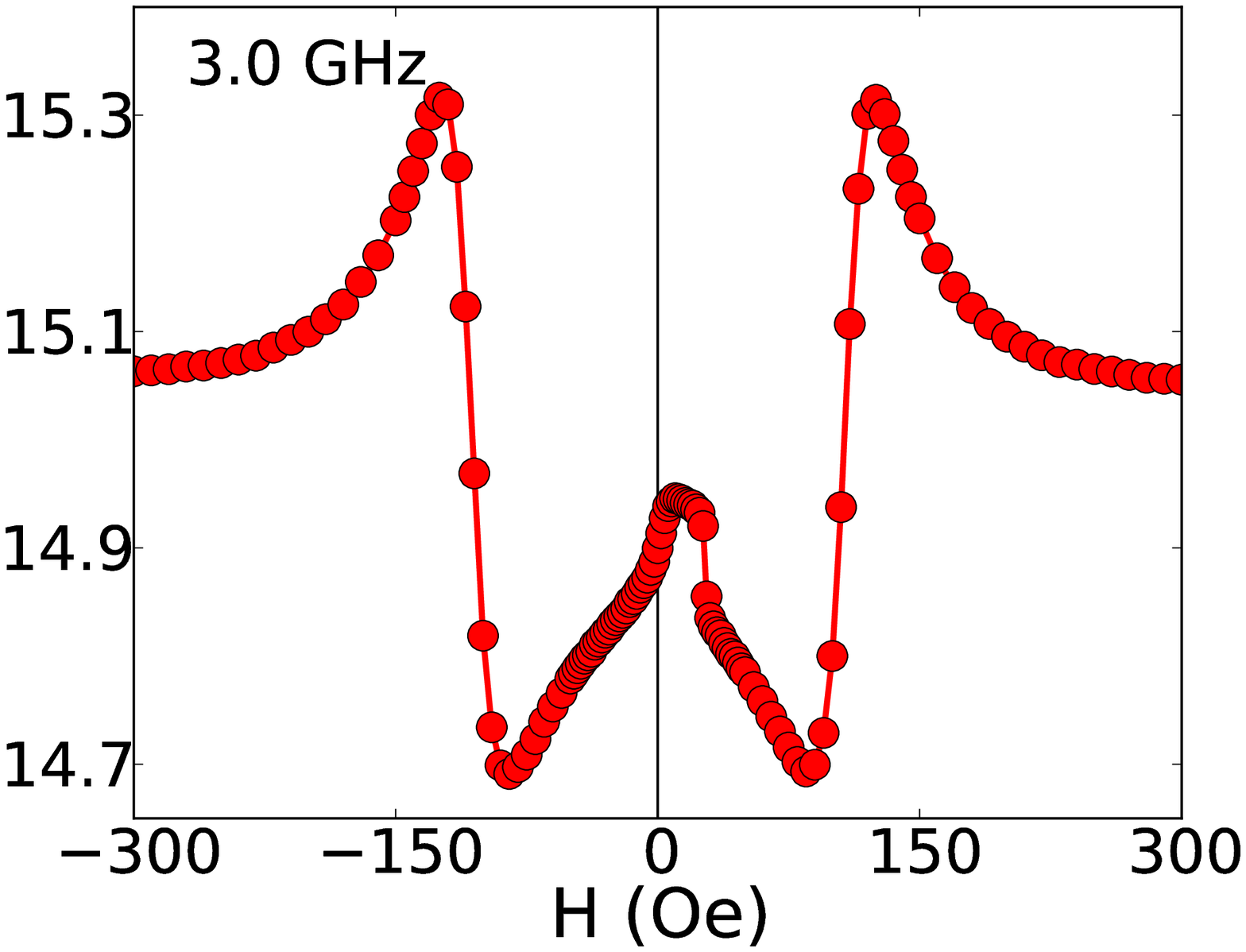}
\end{center}
\vspace{-.3cm}\caption{Evolution of the experimental MI curves for selected frequencies for the ferromagnetic biphase fil with $t_{Cu} = 7$ nm. Similar results are obtained for all the ferromagnetic films with $t_{Cu}$ above $3$ nm and biphase magnetic behavior. We show just part of the curve, when the field goes from negative to positive values. 
} 
    \label{Fig_03} 
\end{figure}

On the other hand, films with $t_{Cu} \geq 3$ nm present noticeable asymmetric magnetoimpedance effect. The asymmetric behavior is assigned by two characteristic features: shift of the MI curve in field, depicted by the asymmetric position of the peaks, and asymmetry in shape, evidenced by the difference of amplitude of the peaks. Figure~\ref{Fig_03} shows the evolution of the MI curves, at selected frequencies between $0.5$ GHz and $3.0$ GHz, for the film with $t_{Cu} = 7$ nm, as an example of the experimental result obtained for the ferromagnetic films with $t_{Cu}$ above $3$ nm. Here, it is important to notice that the presented MI behavior is acquired when the field goes from negative to positive values. However, the MI curves are acquired over a complete magnetization loop and present hysteretic behavior. In particular, when the field goes from positive to negative values, the MI behavior is reversed.

From Fig.~\ref{Fig_03}, regarding the position of the peaks, since the skin effect commands the dynamical behavior, the peaks remain invariable at the low frequency range. For this film, the peak at negative field values is located at $\sim -4$ Oe, while the one at positive fields is at $\sim +30$ Oe. For the other films with $t_{Cu} > 3$ nm and biphase magnetic behavior, the peak at positive field is placed at similar value, although the location of the peak at negative field present dependence with $t_{Cu}$, as will be discussed. With respect to the amplitude of the peaks at low frequencies, for all films with $t_{Cu} > 3$ nm, the MI behavior exhibits an asymmetric two-peak behavior, with the peak at negative field being with higher amplitude than the peak at positive field. As a signature of the emergence of the FMR effect, the displacement of the peak at negative field begins at $\sim 0.6$ GHz, while the position of peak at positive field starts changing at $\sim 1.1$ GHz. Above $ \sim 1.5$ GHz, strong skin and FMR effects are responsible by the MI variations. At this high frequency range, the asymmetry still remains in the portion of the impedance curve around the anisotropy fields. However, the displacement of the peaks toward higher fields suppress the impedance peak asymmetry, in position and amplitude, resulting is symmetric peaks around $H = 0$ with same amplitude. For the film with $t_{Cu} = 3$ nm with intermediate magnetic behavior, similar features are observed respectively at $\sim 0.75$ GHz, $\sim 1.1$ GHz, and $\sim 2.0$ GHz.

\begin{figure}[!] 
    \includegraphics[width=8.5cm]{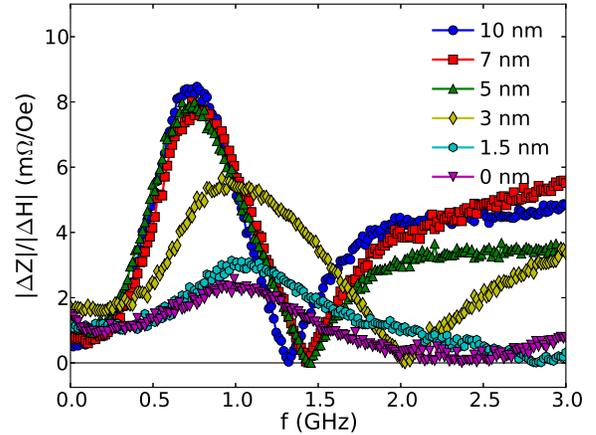}
\vspace{-.3cm}\caption{Frequency spectrum of impedance variations between $\pm 6$ Oe for the films with different thicknesses of the non-magnetic Cu spacer material $t_{Cu}$, indicating the sensitivity around zero field. Notice the kind of saturation effect observed as the $t_{Cu}$ increases above $3$ nm, related to the amplitude and frequency at which the maximum impedance change is reached.}
    \label{Fig_04}
\end{figure}

The most striking finding in the dynamic magnetic response resides in the asymmetry of the MI curves measured for the films with biphase magnetic behavior. It is important to notice that the magnetoimpedance response is nearly linear for low magnetic field values, and the shape of the $Z$ curves depends on the thickness of the non-magnetic Cu spacer material and probe current frequency. As a consequence, the best response can be tuned by playing with both parameters. Figure~\ref{Fig_04} shows the frequency spectrum of impedance variations between $\pm 6$ Oe, as defined by Eq.~(\ref{eq_01}), for each film, indicating the sensitivity around zero field.

From the figure, we verify that the films split in different groups according the sensitivity around zero field. It is important to notice that each one is related to a given magnetic behavior, verified through the magnetization curves. Films with $t_{Cu} < 3$ nm have the largest sensitivity values at $\sim 1.0$ GHz, the film with $t_{Cu} = 3$ nm at $\sim 0.9$ GHz, while the ones with $t_{Cu} > 3$ nm at $\sim 0.75$ GHz. For all of them, the sensitivity peak is found to be at frequencies just after the FMR effect starts appearing, and while the MI peaks are still placed close to the anisotropy fields. The highest sensitivity is observed for the films with $t_{Cu} > 3$ nm, reaching $\sim 8$ m$\Omega$/Oe, and seems to be insensitive to the thickness of the non-magnetic spacer material. In this situation, the AMI curves have nearly linear behavior at low magnetic field values, and the slope of the linear region at nearly zero field is negative, due to the shape of the MI curve.

Our results raise an interesting issue on the behavior of the peaks in the magnetoimpedance effect and the energy terms affecting the transverse magnetic permeability. Generally, our films consist of two ferromagnetic layers, with distinct anisotropy field strengths, intermediated by non-magnetic spacer material. We interpret our experimental data as a result caused by the competition between two types of magnetic interactions between ferromagnetic layers: exchange coupling between touching ferromagnetic phases, and long-range dipolarlike or magnetostatic coupling~\cite{JAP87p5759}. In particular, they are strongly dependent on the thickness of the spacer material, and the action of each one affect in different ways the MI behavior.

If both ferromagnetic layers are quasi-saturated, where there are no walls with wall stray fields, the coupling should adjust the magnetization of the two ferromagnetic layers parallel to each other. For $t_{Cu} < 3$ nm, the strong coupling is caused by the exchange interaction between touching ferromagnetic layers and through pinholes in the non-magnetic spacer, and the whole sample behaves as a single ferromagnetic layer~\cite{TSF520p2173}. In this sense, we confirm the expected symmetric magnetoimpedance behavior of single anisotropic systems. 

For $t_{Cu} > 3$ nm, the Cu layer is completelly filled~\cite{TSF520p2173}, and the nature of the coupling is magnetostatic. If the ferromagnetic layers were completely uncoupled, one could expect multiple peak MI behavior, associated to the anisotropy fields of each different layers, around $\pm 30$ Oe and $\pm 10$ Oe. This behavior is not verified here, indicating that the AMI can not be explained assuming independent reversal of the NiFe and Co layers. Thus, the asymmetry arises as a result of the magnetostatic coupling between the ferromagnetic layers. The origin of the magnetostatic coupling is ascribed to the hard Co magnetic phase in terms of an effective bias field, induced by divergences of magnetization mainly due to roughness in the interfaces and limits of the sample~\cite{JAP105p033911}, that must be taken into account as a contributor to the transverse permeability. The field penetrates the non-magnetic spacer layer and results in a torque on the magnetization of the opposite layer. It is important to point out that the anisotropy field of the hard Co layer is considerable larger than the soft NiFe layer, the reason why this asymmetric behavior is not verified in traditional multilayers.

In this sense, the main features of the asymmetric magnetoimpedance verified in films with biphase magnetic behavior can be explained through the effective interaction between the ferromagnetic layers. The influence on the soft NiFe layer is dependent on the magnetic state of the hard Co layer, as well as on the thickness of the Cu layer spacer. The difference of amplitude of the peaks is understood from the magnetic saturated state in terms of the orientation of the two layers. The peak at negative field is higher than that in positive field, since the magnetization of the soft NiFe layer is parallel to the magnetization of the hard Co layer and consequently to the magnetostatic field, as well as to the external field. Since the magnetization of the NiFe layer is reverted as the field is increased, the sense of this magnetization with respect to the magnetostatic field is modified, and this form closes the magnetic flux, resulting a lower peak~\cite{JAP105p033911}. Similar dependance with the orientation of magnetizations has already been verified in field-annealed Co-based amorphous ribbons~\cite{APL75p2114}. When the MI measurement is analyzed for decreasing magnetic field, the reverted behavior is observed, with the higher and smaller peaks at positive and negative fields, respectively, since the sense of the magnetization of the hard Co layer is the opposite.

\begin{figure}[!] 
    \includegraphics[width=8.5cm]{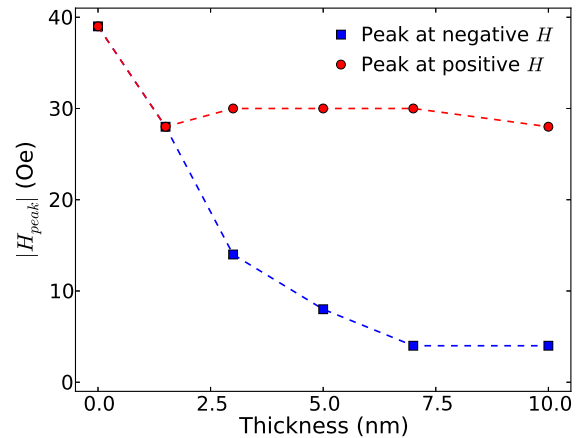}
\vspace{-.3cm}\caption{Magnetic field values in which the impedance peaks are located, at $0.5$ GHz, for the films with different thicknesses of the non-magnetic Cu spacer material $t_{Cu}$. Notice that the films with $t_{Cu} \leq 1.5$ nm present double peak behavior, symmetrical at aroud $H = 0$.}
    \label{Fig_05}
\end{figure}

By employing MI measurements, it is possible to estimate the effective coupling strenght between the NiFe and Co layers. This can be done by considering the location of the peaks in the MI curves at the low frequencies. Figure~\ref{Fig_05} shows the magnetic field values in which the impedance peaks are located, for different thicknesses of the non-magnetic Cu spacer material, at the low frequency range. The position of the impedance peak at negative field presents a noticeable dependence with $t_{Cu}$. In particular, it is verified a reduction of the field value where the peak is located as $t_{Cu}$ is increased, corroborating the assumption of a magnetostatic origin of the coupling between the ferromagnetic layers. In this sense, we interpret the reduction as an indication of the decrease of the bias field intensity {\it acting on} the soft NiFe layer as the Cu thickness is increased. On the other hand, the peak at positive field is located at $\sim 30$ Oe, except for the sample without the non-magnetic spacer material. The constancy in the peak location, irrespective of $t_{Cu}$ and field value of the Co reversion for each sample, suggests that this value corresponds to a intrinsic feature of the ferromagnetic Co layer, since it presents similar thicknesses for all samples. Thereby, we understand it as the magnitude of the bias field {\it induced} by the hard Co layer. 

In conclusion, we have investigated the magnetoimpedance effect in ferromagnetic NiFe/Cu/Co films and observed the dependence of the MI curves, in particular, amplitude and position of the peaks, with the thickness of the non-magnetic Cu spacer material. We have verified that the MI response of these films can be taylored by the kind of magnetic interaction between the ferromagnetic layers. In this sense, the coupling between the layers is usefull to develop materials with asymmetric MI. From the results, we have observed the crossover between two distinct magnetic behavior, associated to distict kind of the magnetic behavior between the ferromagnetic layers, exchange interaction and magnetostatic coupling, as the Cu thickness is altered crossing through $t_{Cu} = 3$ nm. Thus, we have tuned the linear region of the asymmetric magnetoimpedance curves around zero magnetic field by varying the thickness of the non-magnetic spacer material, and probe current frequency. The highest sensitivity is observed for the films with $t_{Cu} > 3$ nm, reaching $\sim 8$ m$\Omega$/Oe, and seems to be insensitive to the thickness of the non-magnetic spacer material. These results extend the possibilities for application of ferromagnetic films with asymmetric magnetoimpedance as probe element for the development of auto-biased linear magnetic field sensors, placing films with biphase magnetic behavior as promissing candidates to optimize the MI performance.

\begin{acknowledgments} 
The authors thank Vivian Montardo Escobar for the fruitful discussions. The research is supported by the Brazilian agencies CNPq (Grants No.~$471302$/$2013$-$9$, No.~$310761$/$2011$-$5$, No.~$555620$/$2010$-$7$), CAPES, and FAPERN (Grant Pronem No.~$03/2012$). M.A.C.\ and F.B.\ acknowledge financial support of the INCT of Space Studies.
\end{acknowledgments} 

\bibliographystyle{References_h-physrev3}
\bibliography{References_MI}

\end{document}